# spotFuzzer: Static Instrument and Fuzzing Windows COTs

**Yeming Gu[1], Hui Shu[1, *], Rongkuan Ma[1], Lin Yan[1] and Lei Zhu[1]**

1 State Key Laboratory of Mathematical Engineering and Advanced Computing, Zhengzhou, 450000, China
*Corresponding Author: Hui Shu. Email: shuhui123@126.com

**Abstract:** The security research on Windows has received little attention in the academic circle. Most of the new methods are usually designed for Linux system, and are difficult to transplant to Windows. Fuzzing for Windows programs always suffering from its closed source. Therefore, we need to find an appropriate way to achieve feedback from Windows programs. To our knowledge, there are no stable and scalable static instrumentation tools for Windows yet, and dynamic tools, such as DynamoRIO, have been criticized for their performance. To make matters worse, dynamic instrumentation tools have very limited usage scenarios and are impotent for many system services or large commercial software. In this paper, we proposed spotInstr, a novel static tool for instrumenting Windows binaries. It is lightweight and can instrument most Windows PE programs in a very short time. At the same time, spotInstr provides a set of filters, which can be used to select instrumentation points or restrict the target regions. Based on these filters, we propose a novel memory-sensitive instrumentation method which can speed up both instrumentation and fuzzing. After that, we design a system called spotFuzzer, which leverage the ability of spotInstr and can fuzz most Windows binaries. We tested spotInstr and spotFuzzer in multiple dimensions to show their superior performance and stability.

**Keywords:** fuzzing; static instrumentation; binary rewrite; windows

## 1 Introduction

Security research and software analysis technologies on Windows cannot match its market share. The focus of academic research remains on UNIX-like platforms. One of the main reasons is that most applications software on Windows are closed source, which requires more effort for researchers to do a lot of reverse engineering. There is no doubt that Windows is the most widely used operating system. We should pay more attention to its software security.

Vulnerabilities are the main threat to system security. Security researchers use static analysis [1] or dynamic analysis to locate vulnerabilities in software. In our experience, one of the most popular dynamic methods for vulnerability mining is Fuzzing [2]. Especially since AFL [3] appeared in 2013, fuzzing has made great progress. We can find fuzzing tools for file parser [4], system kernel [5], net protocol [6] or IoT devices [7]. The feedback technology introduced by AFL is still the most effective way to find vulnerabilities. Over the years, there are a lot of AFL-like tools [8][9][10] have been developed for different scenarios. The key idea of the feedback technology is leveraging the instrumentation technology to trace the execution path. The default instrumentation mode of AFL is to patch the compiler and insert some code snippet into the target. This compile-time instrumentation has minimal side effects on the target, so it is the preferred choice for AFL. To cope with the closed-source software, AFL also supports the QEMU [11] mode, which uses a virtual machine to dynamically trace the execution path of the target. After 3 years of waiting, the Windows version of AFL was finally released in 2016. WinAFL [12] made a lot of changes to adapt to Windows. It uses DynamoRIO [13] to instrument target dynamically instead of QEMU mode, and drops the compile way. Finally, WinAFL implements roughly the same feedback capabilities as AFL.

Although many practical problems have been solved in the field of fuzzing, there are still many shortcomings. AFL and its successors can only be used for Linux platform. WinAFL was designed for Windows, but it uses DynamoRIO, which makes the target much slower and its applicability is limited. DynamoRIO can only run some simple programs with acceptable overhead. If we want to fuzzing COTs on Windows, we need to come up with a new approach to overcome these problems.

In this paper, we have designed a new fuzzing system for Windows. The system relies on static instrumentation against Windows binaries. The key idea of instrumentation is to extract memory points by reverse analysis, and instrument the target at these points using binary rewriting technologies. We find that existing tools always pursue high rate of basic block coverage and instrument as more as they can. According to our experience, most of the regions in a program are vulnerable free. Therefore, it is not a good idea to instrumenting everywhere in a program, which leads to higher overhead for analysis, instrumentation, and execution. We propose a novel method to filter the points of instrumentation, which can make static instrumentation more efficient, lightweight, and robust.

We developed spotInstr as our static instrumentation tool, which can be divided into two parts. The analysis front-end for extracting memory points and the binary rewriting back-end for instrumenting the target. The analysis front-end was designed as a plug-in for IDAPro [14]. It leverages the advantage of IDAPro's disassembly capabilities and uses its interfaces to analyze the target binary. We have done extensive work to understand the Intel instructions [15] to extract the basic blocks in the assembly code. We also implemented a set of interfaces for filtering the memory points. The back-end is based on PeLib [16]. We have few choices for PE file manipulation libraries. After some research, we finally found the PeLib, an old and no longer maintained Library. PeLib is not well developed and still has many bugs in it. So, we made many patches to make it work properly. In the process of instrumenting, we found that both the analysis and instrumentation phases took a lot of time when working on large binaries. We made a lot of optimizations on the algorithms in both stages and achieved significant performance improvements.

We developed spotFuzzer based on spotInstr. The most obvious improvement of spotFuzzer is that it uses a new architecture for fuzzing running processes on Windows. We find that some programs on Windows cannot start directly or always depend on another program, so ordinary fuzzer can't fuzz these targets directly. SpotFuzzer uses an agent to inject into the target process and builds a belt within the target and the fuzzer.

We demonstrate the applicability of spotInstr and spotFuzzer by instrumenting and fuzzing more than 20 COTS software or Windows components. First, we compared spotInstr with pe-afl [17] and syzygy [18]. Then, we compared spotFuzzer with pe-afl and WinAFL. In conclusion, our instrumentation tool runs dramatically fast. Compared to pe-afl, the average time cost of our tool is about 90% lower and the compatibility is better. Thanks to spotInstr, our fuzzing tool also has better performance than pe-afl and WinAFL, and it can find more vulnerabilities in less time.

This paper makes the following contributions:

First, we developed an instrumentation tool for Windows binaries, called spotInstr, which support almost all Windows PE files, and offers a significant performance improvement over the state of art tools.

Then, we designed a new memory-sensitive instrumentation technique that focus on memory-related vulnerabilities. This technique reduces the number of instrumentations significantly and makes the target execution speed close to the original one, while having the better vulnerability discovery capabilities.

We also propose a new fuzzing architecture for Windows components, called spotFuzzer, which leverage the capabilities of spotInstr. SpotFuzzer makes the fuzzing Windows COTs much easier and offers better performance than the popular WinAFL.

## 2 Motivation

### 2.1 instrument binary-only programs on Windows

Most commercial Windows software are binary-only programs, security researchers must instrument these programs statically before fuzzing them.

But at present, there is no stable static instrumentation tools for Windows yet. The main challenge for static instrumentation is how to rewrite a PE file correctly and quickly.

### 2.2 focus on memory issues

Fuzzing with sanitizers is the most effective way to find memory-related vulnerabilities. When fuzzing on Linux, there are several sanitizers to use to detect memory issues, like AddressSanitizer, MemorySanitizer and LeakSanitizer. But one must use the compiler to recompile the target with source code.

Another way to speed up detecting memory issues is called Selective Instrumentation. According to our experience，when we compile a simple program, the compiler will generate hundreds of functions and thousands of basic blocks, while the actual main function only contains several lines. This indicates that there always be a lot of non-functional codes in target binary. If we want to fuzzing a target, we'd better skip these codes or focus on memory-related areas to save instrumentation time and approve fuzzing performance.

### 2.3 fuzz programs on Windows

Most user-end software is GUI-based, which means we can't interact with it through command line. Most fuzzer, like WinAFL, expects the target grogram match the execution pattern: input-parse-exit. But the GUI-based programs cannot be tested automatically without human intervention.

For Windows, there are some programs which can't start directly or rely on other components. For example, most services on Windows corresponding to a DLL file. We can't load that DLL file in a proper way without change its behavior. The best way to load it is starting its service as normal. But neither WinAFL nor pe-afl can fuzz as that way.

In this work, we address all these bottlenecks, and proposed two handy tools for instrumenting and fuzzing Windows binaries.

## 3 Design

As mentioned previously, there is a big gap between theory and practice for static instrumentation for Windows binaries. We can't find any static instrumentation tool for PE64, and the on-hand tools for PE32 is just too old to fit new PE format. By the way, we need implement a set of interfaces for user to control where to instrument. So, we design spotInstr as a lightweight, robust, and efficient static instrumentation tool for both PE32 and PE64.

When the target binary is finish instrumented, we use spotFuzzer to load and test it with the feedback technology. For some Windows service-related target, we design an agent-based fuzzing framework. All these efforts we make are aiming to the goal: fuzzing Windows binaries with static instrumentation easily, scalable, and efficiently.

### 3.1 System Overview

Our fuzzing workflow contains two stages: instrumentation and fuzzing. When a target is chosen for fuzzing, font-end of spotInstr should be used in the first stage to analyze and extract memory points. Then back-end of spotInstr complete the binary instrumentation. In the second stage, spotFuzzer fuzz the instrumented binary instead of the original one. The top view of our system as shown in Fig. 1.

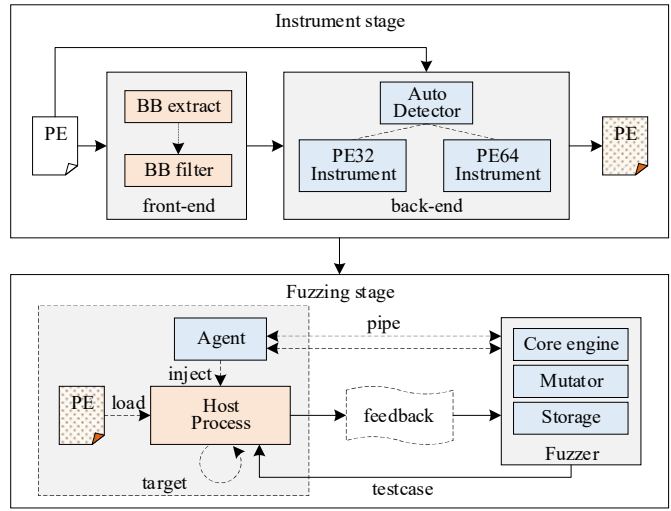

**Figure 1:** Top view of spotFuzzer

*3.2 Basic block extract*

The purpose of static instrumentation is to cooperate with the fuzzer, while spotFuzzer read the feedback of execution path from the instrumented binary. So firstly, we need to find out all basic block in the target binary. The basic block is a sequence of contiguous instructions that contains no jumps or labels, as shown in Fig. 2.

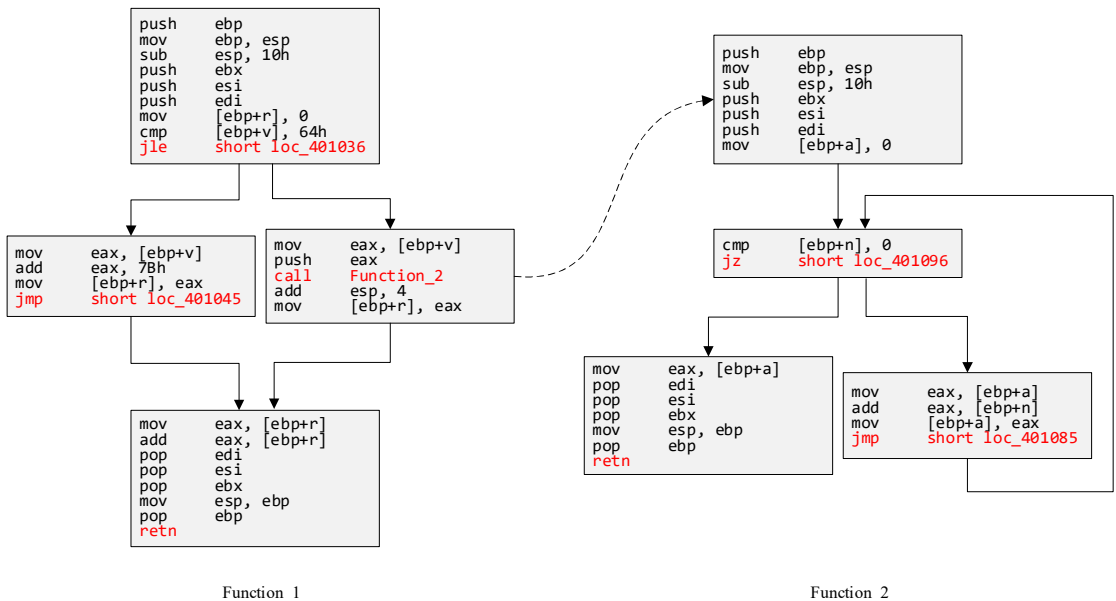

**Figure 2:** Basic blocks in program

It is easy to observe that the basic block always starts at the function entry, the call destination, the jmp destination or the jcc destination. And the basic block always ends with a jmp instruction, a jcc instruction, a ret instruction or ends before next basic block head. To extract all basic block's head, we should analyze all assembly codes in the text segment. The key point is to find all control flow transfer instructions, which include call, jmp, jcc, and ret. And calculate out the destination address according to the operand value.

**CALL Instruction:** The call instruction has 10 different formats according to Intel's Instruction Set

Reference. It is easy to identify such instructions, which always starts with Opcode 0xE8, 0x9A or 0xFF. In 64-bit mode, we should take care of the REX prefixes in the instruction. It is more complex to calculate the destination address of the call instruction. Different calculation method should be taken for different operant type.

**JMP Instruction:** The jmp instruction has 11 different formats according to Intel's Instruction Set Reference. We can find these instructions with Opcode 0xEB, 0xE9, 0xEA or 0xFF. Also, we should consider the REX prefixes in 64-bit mode. The calculation of destination address is similar to the one of call instruction.

**JCC Instructions:** The jcc instructions include a set of conditional jump instructions, which include ja, jb, jc, etc. There are 95 different formats of opcode according to Intel's Instruction Set Reference. Jcc instructions always start with one byte [0x70~0x7F] or two bytes 0x0F + [0x80~0x8F]. The calculation of destination address is similar to the one of call instruction.

**Jump Tables:** The most difficult situation is the special jmp instructions which we call them jump tables. These instructions always use a register as its operand (like jmp rcx). It is hard to calculate the destination address, but we can use context before the jmp instruction to figure out the position of jump table. With the jump table data, we can then calculate all destination addresses. Fig. 3 shows an example of Jump Table.

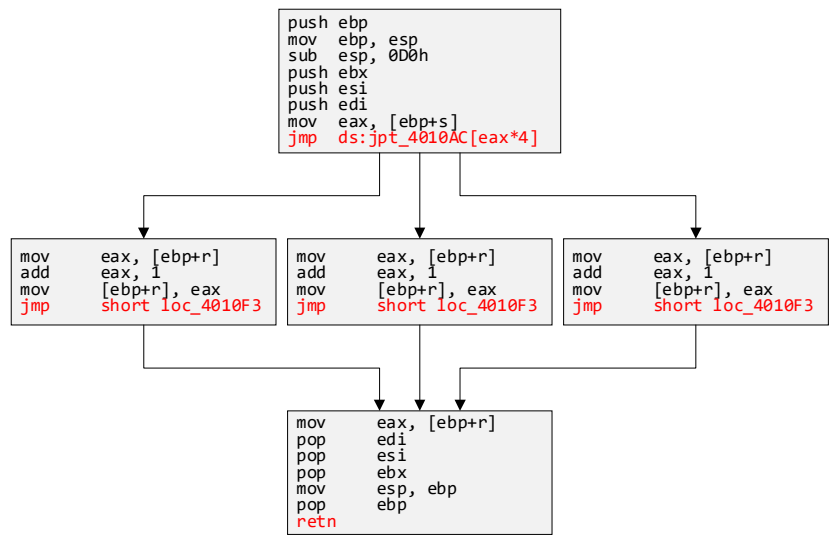

**Figure 3:** Jump Table in program

We design two different algorithms to detect jump tables in PE32 and PE64. Then we can add all destination address in jump tables into the basic block list.

### 3.3 Instrument points filter interfaces

With all the basic block information, we can take some strategies to filter the basic block. According to observation, there are a lot of non-functional code, initializing code and helper code in target. Most of which have little chance to trigger critical vulnerability. Instrumenting on these codes increase the analysis time, instrumentation time, and slow down the execution. Even worse, more instruments also increase the possibility of program errors. In this paper, we provide 3 interfaces for user to include or exclude some basic blocks.

**Address including:** The basic block list contains all basic block information contains starting address, instruction size, relative address position, etc. and the list should be sorted by the starting address. So, it is very simple to specify an address or address range to tag as included and then delete all other items which isn't tagged.

**Address excluding:** The basic block who's address specified to be excluded will be deleted from the basic block list. Because the basic block list is sorted by the starting address, the deleting should be very fast.

**Function name regular matching:** For some cases, we may have the symbol file for the target or just rename a set of functions. Then we can filter the basic block list by the function names. At this moment, spotInstr support using regular expression to include or exclude functions, basic blocks in which will be included or excluded.

*3.4 Static binary rewriting for PE file*

In this paper, we support both trampoline and inline mode to instrument code snips. The trampoline technology to realize static binary rewriting, which means a 5-bytes jump instruction will replace origin codes and redirect the control flow to a trampoline. This technology has obvious advantages: simpler, faster, more stable, more reliable, and lightweight. Fig. 4 shows the PE file structs for non-instrumented and instrumented binaries.

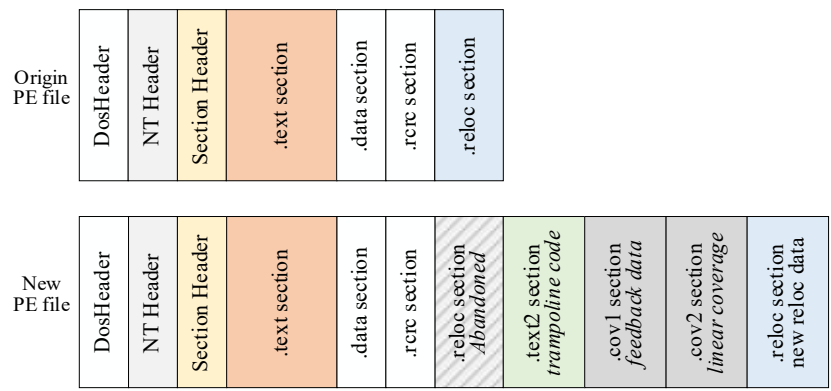

**Figure 4:** Windows PE file struct

**PE structure (32bit or 64bit) auto detect:** The spotInstr back-end support both PE32 and PE64, and there is no need for user intervention. In order to realize this detector, we build a PE file parser for both PE32 and PE64, with which the spotInstr can recognize the PE structure before doing any instrument. This work greatly improves the usability of the tool.

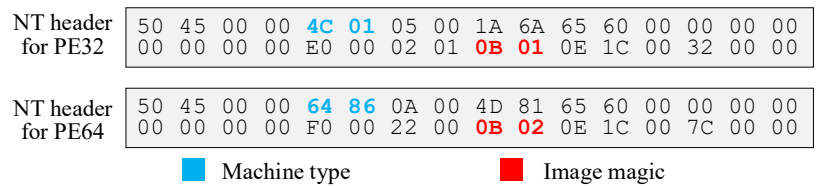

**Figure 5:** Flags in NT header for PE32 and PE64

**Build trampoline segment:** The trampoline segment is used to store all the trampoline code snippets. According to our implementation, each memory point should have its own trampoline code snippet. The size of this segment should be calculated according to account of memory points and the flag of this segment should be set to EXECUTE_READ.

**Build feedback segment:** The feedback segment is used to store feedback data (e.g., execution path bitmap), which will be used by the fuzzer. In this paper, we inherit the feedback data structure from WinAFL. In addition to this, the feedback segment also holds a size field which indicates the size of the extra feedback segment. We use the extra feedback segment for records linear basic block coverage information when user turns it on. The linear basic block coverage information can be used for lighthouse in IDAPro.

**Build local storage segment:** The local storage segment or TLS segment is used to isolate storage between threads. That means each thread will maintain a TLS segment for local data storage. We use this segment to hold the last basic block address and the jump back address for resume the origin control flow. So, even if the target is multi-thread program, the execution paths for each thread won't be confused.

**Update the relocation table:** The relocation table is very important for PE file to calculate the right addresses. The instrumentation moves the original code to trampoline which make the original relocation information is no longer correct. After all memory points have been processed by the spotInstr, the relocation table should be updated to fix all relative addresses in trampolines. The trampolines locate in the new segment, that means the virtual address may exceed the relocation table. So, the simplest way to correct the relocation table is to add some new entries at the end of the table. The old entries for addresses in replaced instructions must be deleted to avoid relocation breaking the jump instructions. In summary, updating the relocation table should have 2 processing stages: the cleaning stage and the inserting stage.

**Update global fields and checksum:** After all the processes above have been finished, we should update some global fields in PE header, such as BaseRelocRva, BaseRelocSize, etc. Before updating the checksum, the old one must be reset to 0.

### 3.5 Fuzzing framework with static instrument

The latest version of WinAFL support instrumenting a binary via syzygy statically, but syzygy only provides a framework able to decompose PE32 binaries with full PDB. That is useless for most COTS software, even the Windows components rarely have a private symbols file. So, we have to abandon syzygy and replaced it with our spotInstr.

**General fuzzing:** If the target binary can be loaded normally, spotFuzzer will act just like WinAFL to fuzz the target. It's a gool idea to write a harness to load a DLL file as used in WinAFL. This general fuzzing should be suitable for most software. Fig. 6 shows the general fuzzing framework.

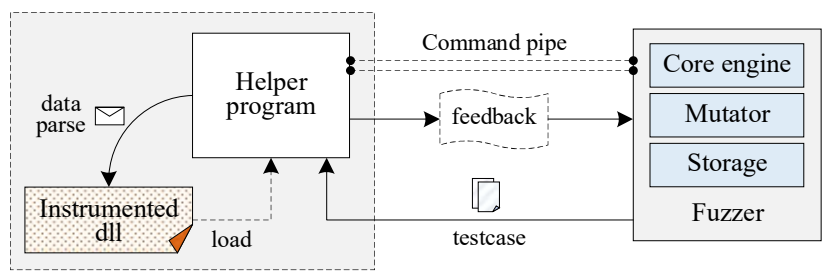

**Figure 6:** Framework of general fuzzing

**Agent based fuzzing:** If the target is a service on Windows or can't be loaded normally, spotFuzzer will inject an agent into the target process. The agent use named pipe to communicate with spotFuzzer, and once injected it will register an exception handler to catch crash information. Fig. 7 shows the agent based fuzzing framework.

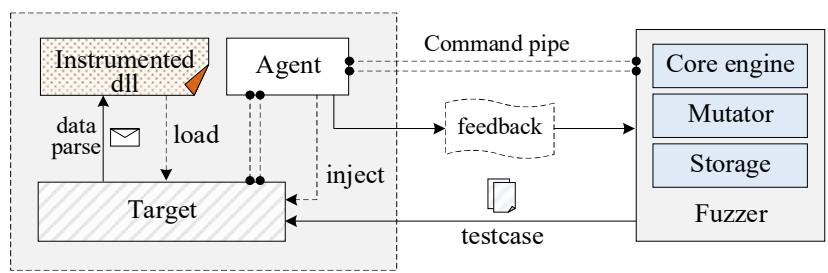

**Figure 7:** Framework of Agent based fuzzing

## 4 Evaluation

In this section we evaluate scalability of spotInstr, speed, and overhead compare to pe-afl. We also evaluate performance of spotFuzzer on some Windows COTS software.

### 4.1 Instrument Scalability

We evaluate spotInstr on several widely used software on Windows, such as 7z, notepad++, WinRAR and 010editor. We also choose some system component additionally. Tab. 1 shows a list of all successfully instrumented binaries on Windows.

**Table 1:** Binaries list for scalability test

| Binary | Vender | Architecture | spotInstr | pe-afl | syzygy |
|---|---|---|---|---|---|
| 7za.exe | 7-Zip | PE32 | ✓ | ✓ | ✗ |
| notepad++.exe | Notepad++ | PE32 | ✓ | ✗ (crash) | ✗ |
| rar.exe | RarLab | PE32 | ✓ | ✗ (crash) | ✗ |
| 010Editor.exe | SweetScape | PE32 | ✓ | ✗ (crash) | ✗ |
| cmake.exe | CMake | PE32 | ✓ | ✓ | ✗ |
| jscript.dll | Windows | PE32 | ✓ | ✓ | ✗ |
| imagingengine.dll | Windows | PE32 | ✓ | ✓ | ✗ |
| gdi32.dll | Windows | PE32 | ✓ | ✓ | ✗ |
| mpengine.dll | Windows | PE32 | ✓ | ✗ (failed) | ✗ |

The results show that our tool can correctly instrument all these executable program or dynamic libraries, while pe-afl can work on a part of them and syzygy can instrument none of them. The main problem for pe-afl. Syzygy need private pdb file for the target. That means syzygy only support targets with source code, one can recompile it and generate the pdb file.

Besides, we also compare some usable features, such as instrument mode, target architecture, thread mode, pdb file dependence, and selective instrumentation. Jump mode is more light weight than inline mode, it will be more stable and efficiency to instrument a huge target with selective Instrumentation. For programs that contain only one parsing thread, single-thread mode can reduce runtime overhead significantly than multi-thread mode. Selective Instrumentation make researchers able to focus on more interesting areas. Tab. 2 shows the features supported by spotInstr and other tools.

**Table 2:** Instrumentation features

| Binary | spotInstr | pe-afl | syzygy |
|---|---|---|---|
| Inline mode | ✓ | ✓ | ✓ |
| Jump mode | ✓ | ✗ | ✗ |
| Support 32bit | ✓ | ✓ | ✓ |
| Support 64bit | ✓ | ✗ | ✗ |
| Single-thread mode | ✓ | ✓ | ✗ |
| Multi-thread mode | ✓ | ✓ | ✓ |
| PDB file independent | ✓ | ✓ | ✗ |
| Selective Instrumentation | ✓ | ✗ | ✗ |

### 4.2 instrumentation performance

We try to compare spotInstr with other tools, such as pe-afl and syzygy. As mentioned before, pe-afl support a part of PE32 binaries and syzygy only support PE32 binaries with private symbols. We can hardly find COTS software that meets the requirements. We have to remove syzygy from the performance evaluation. In order to make the comparison more meaningful, we only choose PE32 binaries which can be both instrumented successfully by spotInstr and pe-afl. Tab. 3 shows the binaries chosen for testing. The smallest one is archive.dll with about 176KB, and the biggest is mpengine.dll with about 11MB.

Table 3: Binaries list for performance test

| Binary | Size (KB) | Architecture | GUI | Description |
|---|---|---|---|---|
| archive.dll | 176 | PE32 | No | Library for libarchive |
| 7za.dll | 269 | PE32 | No | Library for 7-zip |
| gdi32.dll | 304 | PE32 | No | Library for Windows GDI |
| eqnedit32.exe | 524 | PE32 | Yes | Formula editor used by MS Word |
| rar.exe | 568 | PE32 | No | Command line tool for WinRAR |
| jscript.dll | 670 | PE32 | No | Microsoft javascript engine |
| 7za.exe | 723 | PE32 | No | Command line tool for 7-zip |
| imagingengine.dll | 1810 | PE32 | No | Microsoft image engine |
| winrar.exe | 2433 | PE32 | Yes | WinRAR GUI program |
| notepad++.exe | 3005 | PE32 | Yes | Notepad++ GUI program |
| jscript9.dll | 3779 | PE32 | No | Microsoft javascript engine |
| cmake.exe | 7865 | PE32 | No | Command line tool for CMake |
| mpengine.dll | 11281 | PE32 | No | Microsoft malware protection engine |

To evaluate the instrumentation performance, we design three tests: output size, time cost and execution overhead. Before testing, we first measure the size of each binary and write a plugin for IDAPro to calculate the number of basic blocks in each binary. Tab. 2 also lists the basic block counts of PE binaries chosen for testing. The smallest one named archive.dll, which has less than 9,000 basic blocks. The mpengine.dll is the core engine of Microsoft Malware Protection service, which is the biggest and contains more than 590,000 basic blocks.

First, we compare the count of basic blocks instrumented by spotInstr and pe-afl. Tab. 4 shows the count of basic blocks instrumented by different tools. On all test programs, spotInstr can instrument about 20% more basic blocks with its inline mode than pe-afl. But for jump mode, spotInstr instruments a little fewer basic block than pe-afl. That's because jump mode only supports a basic block with more than 5 bytes to hold the jump instruction.

Table 4: Basic blocks instrumented by different tools

| Binary | Basic Blocks count | | | | |
|---|---|---|---|---|---|
| | pe-afl | spotInstr | -select | -jump | -jump-select |
|---|---|---|---|---|---|
| archive.dll | 7660 | 8652(↑13%) | 880 | 6645 | 826 |
| 7za.dll | 9452 | 11951(↑26%) | 2117 | 8454 | 1953 |
| gdi32.dll | 14003 | 16781(↑20%) | 1202 | 12569 | 1154 |
| eqnedit32.exe | 10040 | 14448(↑44%) | 947 | 12255 | 920 |
| rar.exe | 21744 | 25455(↑17%) | 2816 | 18731 | 2701 |
| jscript.dll | 29444 | 34988(↑19%) | 4964 | 27421 | 4856 |
| 7za.exe | 30946 | 38326(↑24%) | 6531 | 26892 | 6137 |
| imagingengine.dll | 70224 | 81110(↑16%) | 8738 | 62914 | 8479 |
| winrar.exe | 54317 | 64553(↑19%) | 7311 | 48382 | 6972 |
| notepad++.exe | 64147 | 82095(↑28%) | 7594 | 61113 | 7214 |
| jscript9.dll | 138731 | 172194(↑24%) | 26519 | 130506 | 24301 |
| cmake.exe | 261180 | 312399(↑20%) | 28683 | 234899 | 27510 |
| mpengine.dll | 346443 | 594481(↑72%) | 43284 | 315238 | 40766 |

Second, we compare the size of the output binaries of spotInstr and pe-afl. We use spotInstr and pe-afl to instrument all these programs with their default setting and collect the size of instructed binaries. Fig. 8 shows the sizes of programs instrumented by different tools or with different mode. We find that our tool did a little better than pe-afl on most binaries. While working on some small binaries, spotInstr and pe-afl performed almost the same. We find that different instrumentation modes have a significant impact on the size of the instrumented file. Jump mode always generates smaller output files than inline mode, and the average size reduction is about 10%. Compare to the raw inline mode, if we turn on selective instrumentation, the average size reduction is about 42%. For jump mode, the same selective instrumentation will cause the reduction to 57%.

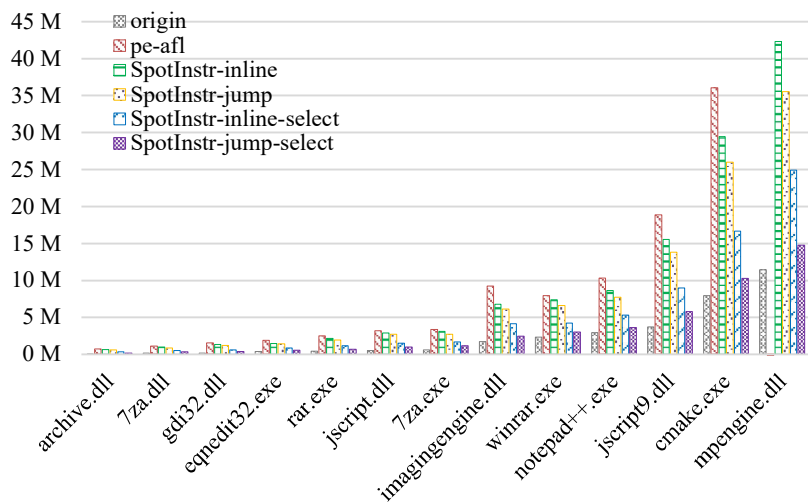

**Figure 8:** Size of original and instrumented binaries

Third, we compare the efficiency of the tools. Fig. 9 shows the time of instrumentation spent by spotInstr and pe-afl. Obviously, spotInstr spend much less time than pe-afl in all tests. Pe-afl took 5x~10x more time than spotInstr on some small binaries, such as archive.dll, 7za.dll, gdi32.dll and eqnedit32.exe. Pe-afl took 30x~100x more time than spotInstr on some bigger binaries, such as rar.exe, jscrip.dll, 7za.exe and cmake.exe, etc. It is worth noting that pe-afl spend more than 1 hour and finally result in an error when instrumenting on mpengine.dll.

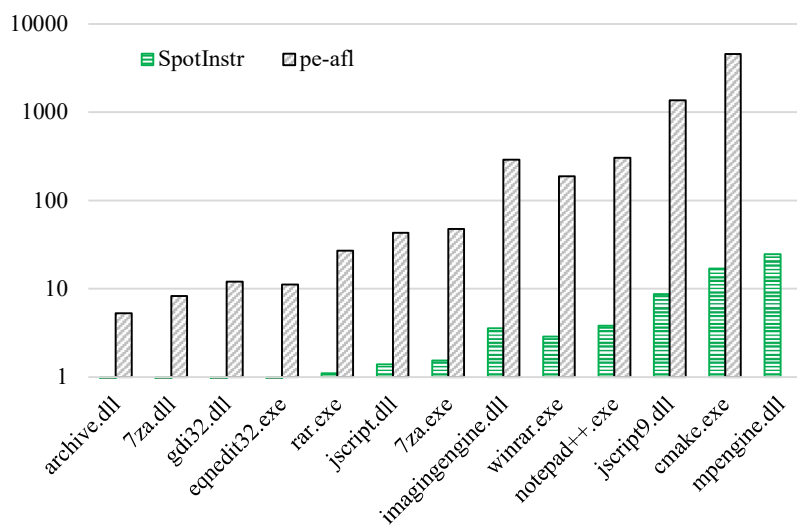

**Figure 9:** Time cost of instrumentation

At last, we compare the execution time between original programs, static instrumented ones, and dynamic instrumented ones. To figure out the execution overhead caused by instrumenting, we select some typical programs for testing. For the convenience of comparison, we try to make the baseline parsing time of input data close to each other. So, we choose the appropriate input data to feed to the instrumented software. In this test, we also add DynamoRIO to show the dynamic instrumentation's overhead. Fig. 10 shows the average execution time of 10 runs with different instrumentation type. According to the result, the overhead of the static instrumentation is much less than the dynamic one. Specifically, the average execution overhead of spotInstr-inline is about 17%, while the overhead of pe-afl is about 13%. The main reason for this small gap is that spotInstr-inline instrument more basic blocks than pe-afl. When we use selective instrumentation, as shown by spotInstr-inline-select in the picture, the overhead reduces to 2.6%.

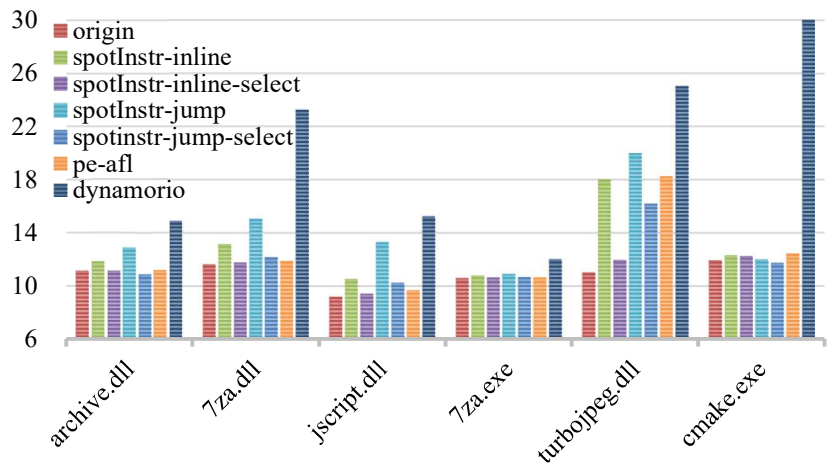

**Figure 10:** Execution time with different instrumentation type

### 4.3 Fuzzing performance

We measured the fuzzing performance from three aspects, including fuzzing speed, execution paths and unique crashes. The WinAFL was used as our baseline method. The spotFuzzer leverage spotInstr to make instrumentation on target program. To look closely at the effect of different instrumentation options on fuzzing, we built spotFuzzer with different instrumentation modes. As a result, we got 4 tools, namely spotFuzzer-inline, spotFuzzer-inline-select, spotFuzzer-jump, and spotFuzzer-jump-select. In which, "inline" and "jump" stand for the instrumentation modes, "select" means the tool uses memory-sensitive technology.

As the use of WinAFL with syzygy is limited, it can't work on most COTS software. We use the dynamic mode for WinAFL instead. In order to test WinAFL, we choose 7za.dll as the fuzzing target, which can run correctly under all these fuzzers.

To compare the fuzzing speed between the fuzzers, we observe the number of execution samples within a certain period. Fig. 11 shows the total samples tested over time and fuzzing speed for spotFuzzer and WinAFL. There is no doubt that all the static instrumentation methods have better performance than WinAFL. We can see that target instrumented with inline mode run much faster than jump mode as expected, which is because jump mode introduces a large number of additional call instructions. However, the interesting thing is that the combination of jump mode and memory-sensitive instrumentation makes the target surprisingly fast.

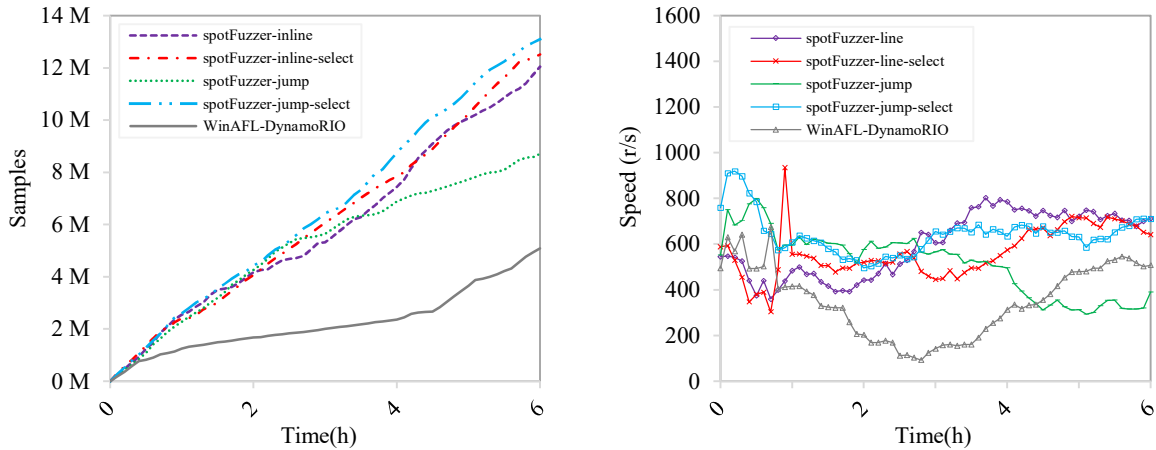

**Figure 11:** Number of total samples and fuzzing speed

Fig. 12 shows the total paths and unique crashes discovered by the fuzzers. As we can see, all spotFuzzers discovered more paths than WinAFL profit from its high instrumentation rate and fast fuzzing speed. Not surprisingly, tools use selective instrumentation discover less paths than full instrumentaion, that is because fewer basic blocks means fewer paths.

The most important performance for a fuzzer is its ability to discover vulnerabilities. Fig. 12 shows that all spotFuzzers find more crashes than WinAFL, especially in the early time of fuzzing.

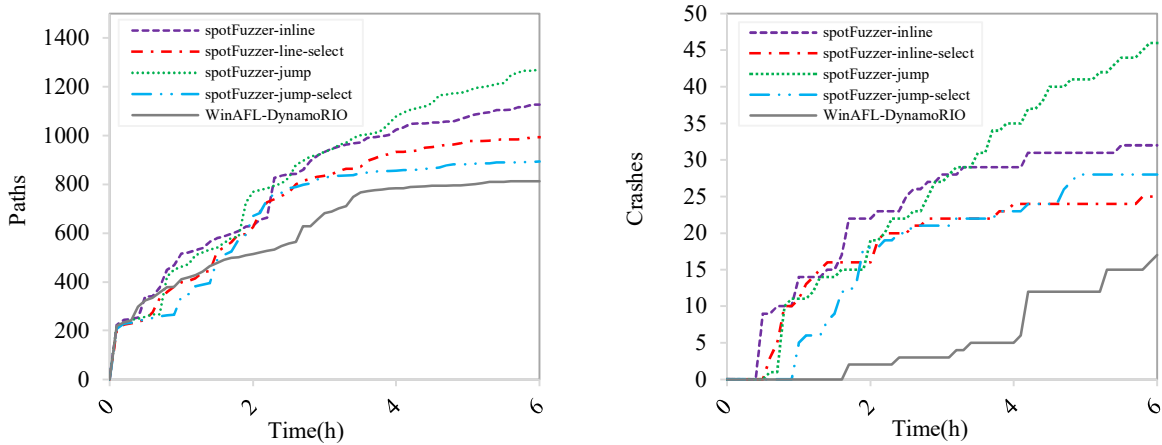

**Figure 12:** Number of total paths and unique crashes

The unique crashes not equal to unique vulnerabilities. After some analyze, we find that a high proportion of the unique crashes cause by the same vulnerable. This problem becomes more serious when more basic blocks selected for instrumentation. As shown in Tab. 5, all fuzzers can find a lot of unique crashes, but among which only a few are unique vulnerabilities. Despite this, our tools found more unique crashes than WinAFL.

Table 5: Unique Crashes vs. Unique Vulnerabilities

| Target | spotFuzzer | | | | WinAFL |
|---|---|---|---|---|---|
| | inline | inline+select | jump | jump+select | |
| Instrument BBs | 11951 | 2117 | 8454 | 1953 | - |
| Execution paths | 1127 | 993 | 1271 | 894 | 813 |
| Unique crashes | 32 | 25 | 46 | 28 | 17 |
| Unique vulns | 3 | 4 | 3 | 4 | 2 |

## 5 Discussion

**Instrument basic block coverage:** In this paper, when spotInstr work on trampoline mode, a 5-bytes jump instruction is chosen to fill the memory point for instrumentation, which means the points contains room less than 5 bytes won't be instrumented. According to our test data, we find such points will less than 10% of the total. We use the neighbor instruction to expand the point's room, which alleviates the problem to a certain degree. But there are still several ones left and can't be instrumented. We notice that e9patch [19] try to reuse the instruction's origin bytes to construct a valid jump instruction. But that may cause a high virtual memory usage and make the process of instrumentation much more complex.

**Static instrumentation on Windows kernel:** The static instrumentation technology introduced in this paper should work with all Windows binaries. In theory, spotInstr can instrument the Windows kernels, with the help of which, researchers can analyze or fuzzing the Windows kernel more efficiently.

**Expand the usage of static instrument:** Lots of researches [20][21] have indicated that program instrumentation plays a very important role in program analysis. Program instrumentation can be used to memory access analysis [22], program behavior analysis [23], data structure recovery [24], and vulnerability mining [25], etc. But most of them are target to Linux or open-source software, there is little research on Windows binaries. We believe that a simple, stable, and usable static instrumentation tool is a good start for Windows binary analysis.

## 6 Related works

In this section, we discuss the related work that are both complementary and orthogonal to our efforts in binary rewriting and fuzzing.

**Binary rewriting:** The rewriting technology can be traced back to the 1990s. At that time, the binary rewriting was mainly used to analyze or optimize the performance of programs, and almost all the tools like ATOM [26], QPT [27], EEL [28] and Etch [29] relied on static rewriting. After 2000, dynamic rewriting has become the mainstream research direction. A lot of successful tools appeared one by one: Dyninst [30], Vulcan [31], Vulgrind [32], DynamoRIO, PIN [33], QEMU, etc. Static rewriting has become a hot research direction again since 2010. At that time, new technology like reassembling was used to regenerate a binary. Tools like PEBIL [34], SecondWrite [35], BISTRO [36], Uroboros [37], Ramblr [38], Multiverse [39], RetroWrite, E9Patch, etc. did a lot of work in theory and observation of static rewriting. Throughout all the static rewriting tools, we find that most of them are for Linux or Unix-like system. Only Etch are designed for Windows binaries, but it is too old for the modem operating system.

**Fuzzing:** Fuzzing is currently the most popular vulnerability discovery technique. Fuzzing was first proposed by Barton Miller at the University of Wisconsin in 1990s [40]. We find that AFL made Coverage-based grey-box fuzzing so popular and almost created a new fuzzing area. Lots of fuzzing tools developed upon AFL like AFL++ [41], AFLGo, AFLPIN [42], AFLSmart [43], FastAFLGo, etc. But for Windows the picture was very different: AFL first released in 2013, while WinAFL released in 2016. WinAFL use DynamoRIO to fetch feedback during execution, and its static mode based on syzygy almost unusable. Then pe-afl was released for fuzzing Windows binaries, but only for PE32. In our experiments, pe-afl may cause some problem errors and made the target crash abnormally. We can hardly find a tool that can fuzzing Windows COTS software with static instrument.

## 7 Conclusion

In this paper, we design two handy tools for instrumentation and fuzzing Windows binaries. The spotInstr is a static instrumentation tool for Windows binaries without source code. It provides trampoline and inline mode for different usage scenario, and supported both PE32 and PE64. In other words, spotInstr can instrument almost any binary on Windows. The spotFuzzer was designed for fuzzing Windows COTS software. For general program, spotFuzzer provides general fuzzing mode just like WinAFL. But for abnormal targets, like system service or kernel module, spotFuzzer can switch to agent mode, and inject an agent to the target for fuzzing. What's more, we develop a memory-sensitive instrumentation method for spotInstr, which can reduce execution overhead and locate vulnerabilities faster.

**Acknowledgement:** We thank the anonymous reviewers for their insightful comments on our work. This paper is supported by the National Key Research and Development Project (2019QY1305). Any opinions, findings, and conclusions or recommendations expressed in this paper are those of the authors and do not necessarily reflect the views of the funding agencies.

**Funding Statement:** The authors received no specific funding for this study.

**Conflicts of Interest:** The authors declare that they have no conflicts of interest to report regarding the present study.